\documentclass[11pt]{article}  
\usepackage{menuproc}
\usepackage{cite}
\usepackage{epsfig}
\usepackage{amsmath,amssymb}
\begin{document}
%
%
%
\titlematter{Meson-Nucleon Physics:  Past, Present and Future}%
{B.M.K.~Nefkens}%
{UCLA, Los Angeles, CA 90095, U.S.A.}%

\begin{center} Abstract\end{center}%
%


We will present some thoughts on the following topics:

1. Major highlights in the history of strong interactions such as
isospin, the pion, SU(3), quarks, the color degree of freedom, QCD.

2. Topics of high current interest such as quark confinement, the 
origin of mass, the
search for the gluon degrees of freedom, chiral symmetry, flavor symmetry,
regularities in the properties of the light-baryon families (parity 
doublets, clusters)
decay patterns, hadronization, chiral restoration, effective 
Lagrangians and their
degrees of freedom.

3. The input parameters for QCD and for effective models.

4. Hadron physics as a necessary aspect of precision tests of the 
Standard Model and
of the Search for New Physics.



\section{Introduction}

Some of the basic questions that have been discussed at MENU-IX, our 
highly successful
symposium which we are concluding this afternoon, are:

\begin{flushleft}
\lq\lq Where are we in the study of Hadron Physics?"

\lq\lq What has been accomplished recently in this field?"

\lq\lq Which are the most interesting problems?"
\end{flushleft}
I will endeavor to address these questions.  In preparing this report 
I have been aided by
the White Papers of two recent workshops, \lq\lq Key Issues in Hadron 
Physics" [1], and
the DNP meeting on Hadronic Physics [2], and two reports on the 
subject by James Bjorken
[3,4].

We are at a momentous time in the short history of Hadronic Physics. 
There is a growing
appreciation for the intellectual challenges presented by QCD in the 
non-perturbative
regime, for the crucial role which it plays in precision tests of the 
Standard Model
(SM), and in the search for New Physics beyond the SM.  There is also 
a growing concern
about the lack of a major laboratory for hadron physics with 
secondary beams.  The
practitioners of Hadron Physics are only loosely organized and they 
are seriously
underfunded.

Our field of physics may be defined as follows [3]:
\lq\lq Hadron Physics is the physics of the hadronic structures and 
the strong -
interacting vacuum".  It is a subfield of quantum chromodynamics. QCD 
at short distance,
$d\sim< 0.1$fm, is a perturbative theory of pointlike quarks and gluons.  Its
validity in this regime is well established, e.g. the strong coupling 
constant $\alpha_s$
has been determined in more than a dozen different ways [5] to an 
accuracy of a few
percent; we also point to the very successful results coming from $Z$ 
and $W$ studies.  At
large distances such as $> 2$ fm, Hadron Physics is a theory of pions 
and nucleons and
their strange counterparts.  It is characterized by the spontaneous 
symmetry breaking of
the approximate chiral symmetry of QCD.  At intermediate distance occur a rich
variety of phenomena including hadron resonances, Regge trajectories, 
soft diffractions and
hadronization.  Except for a few simple lattice - gauge calculations, 
the role of QCD is
very limited; it mainly justifies the use of chiral symmetry, chiral 
perturbation theory,
the $N_c^{-1}$ expansion, and so forth.  The real need for 
calculations of practical
quantities is at intermediate energy.  The mass of the proton and 
neutron have been
measured to an accuracy of a few parts in $10^{7}$.  The best result 
of a recent
lattice-gauge calculation [6] for the proton mass in 878$\pm$25 MeV. 
The hadronic
corrections are now the main limitation to such interesting 
Standard-Model tests as the
new muon anomalous magnetic moment or $g-2$ experiment [7], the 
measurement of CP
violation in the $2\pi^0$ and $\pi^+\pi^-$ decay of $K_L/K_S$ (the so called
$\epsilon^\prime/\epsilon$ experiment) [8,9], the unitarity test of 
the CKM matrix [10] and
precision determinations of $\alpha_s$.  At this MENU-IX symposium 
much of the full
spectrum of non-perturbative QCD has been under the microscope.

\section{Highlights from the Rich History of Hadron Physics}
It is all together appropriate to have a short recitation of major 
highlights of
our field.  They have led to the current models and effective 
Lagrangians of strong
interaction physics, in particular, they led to the creation of our 
major theory, Quantum
Chromodynamics, QCD.

a) One of the oldest concepts in nuclear physics is that of isospin. 
Its role in the
development of the theory of strong interactions is hard to 
overstate.  Isospin was an
early attempt to formulate the universality of the strong interaction 
at a time when
only the neutron and proton had been identified.  Isospin paved the 
way for Yang-Mills
fields and it helped to usher non-abelian theory into strong interaction
physics.  Isospin was one of the earliest cases of a broken symmetry. 
Now, at  the
beginning of the 21st century, we know the origin of isospin 
breaking: it is the
difference in the masses and the electric charges of the up and down 
quarks.  The study of
isospin breaking occurring in the baryons and mesons of different spin and
parity continues to be of great importance.  It is the only way to 
determine the up-down
quark mass difference in various hadronic environments [11].  Isospin 
violation leads to
meson mixing for instance $\pi^0 -\eta$ and
$\rho - \omega$, also to baryon mixing such as $\Lambda -\Sigma^0$.  The modern
perspective is that isospin is a subgroup of a larger symmetry, the flavor symmetry 
of
massless QCD.

b) The pion has had and still has a major impact on strong interaction physics,
especially on phenomenological models for nucleon-nucleon scattering. 
The current
viewpoint is that the pion is the lightest of the 3 Goldstone bosons that are
associated with the spontaneous breaking of chiral symmetry of a massless QCD.

c) SU(3) symmetry was born out of the wealth of great baryon and 
meson spectroscopic
data.  Measurements of SU(3) breaking, for instance  the mass 
difference of the baryon
octet and decuplet ground states, are the sole way for obtaining the 
$s-d$ quark-mass
difference.  SU(3) breaking is responsible for multiplet impurity such as
due to $\eta-\eta^\prime$ mixing.  SU(3) is a special subgroup of the 
(broken) flavor
symmetry of QCD.

d)  Perhaps the most astonishing discovery, which was prompted by abundant new
spectroscopic data, was the quark
being the ultimate hadronic building block.  The quark has several
spectacular properties such as a non-integral electric charge and 
baryon number.  A single,
free quark cannot be observed because it is captive to the asymptotic freedom
condition of QCD.

e)  The latest \lq\lq quantum number" to be discovered is the color 
charge.  Again
it was the well established details of baryon spectroscopy, in this 
case the features
of the $\Delta^{++}$ (1232) resonance, together with our faith in the 
validity of
Fermi-Dirac statistics which brought this about.

f) QCD is now considered to be the theory of all strong interactions. 
It has many
virtues but its major shortcoming is that it cannot be solved in the 
non-perturbative
regime. Thus, it is helpless when one needs to calculate detailed 
properties of the
proton, neutron and complex nuclei.

\section{High-Profile Subjects}
Problems and topics in hadron physics which are drawing considerable 
attention these
days include the following.

a) What is the mechanism that is responsible for the confinement of 
the quarks?  So far
no one has succeeded in deriving the confinement conditions from QCD. 
This is a major
theoretical challenge.  Help could be provided by the determination 
of the regularities
that characterize the confined quark systems.  Parity doublets 
dominate the $N^*$ spectra
in the mass range 1600-2300 MeV [12]. Do they occur also for heavier 
masses?  Parity
doublets are also seen in in the $\Delta,\Lambda,\Sigma$ families. 
There is no evidence
for parity doublets among the three light-quark meson families. 
Baryons above a certain
mass appear to come in clusters as well.  There is currently 
insufficient data to see if
clusters extend to the $N^*$ mass region above 2300 MeV.

An interesting analogy can be made to the confinement of the electron 
in the hydrogen
atom.  In this case confinement does not follow from Maxwell's 
equations which govern
the electromagnetic interactions.  Rather, confinement in the case of 
the hydrogen atom is
the consequence of the quantization conditions from quantum mechanics.  The
major breakthrough to solving this historic confinement puzzle came 
from the interpretation
of the experimentally discovered regularities in the frequencies of 
the spectral lines
of the hydrogen atom known as the Rydberg formula.  Maybe the spectra 
of the excited
states of the light baryons could play the same role in helping to 
understand quark
confinement.

b) What is the origin of the mass of the proton?  Over 99\% of the 
rest mass of everyone
in this room and elsewhere is due to hadronic matter in the form of 
free and bound
protons and neutrons  The proton consists of two up and one down 
quark for the grand
total of 18 MeV and similarly for the neutron.  Where does the 
remaining 98\% come from?
The missing piece is called the quark condensate, it comes from the 
interaction of the
quarks with the vacuum.  This brings up the question: \lq\lq  What 
really is the vacuum?"

c) A fascinating subject is the proposed existence of new forms of 
matter:  hybrids,
glueballs, pentaquarks, bound states of mesons and baryons, 
molecular-type states, etc.
There is a lot of speculation about this subject.  Our main theory is QCD, the
theory of interacting quarks and gluons, so we might expect that 
confined states have
quark and gluon degrees of freedom.  There is no example of a 
certified hybrid baryon, not
even a single good candidate [12].  The established baryons all obey 
SU(3) flavor symmetry
which implies that they are $qqq$ states.  There are a few candidates 
for a glueball
and hybrid in the meson sector. Since there is no free meson target 
available a unique
determination of the spin and parity of a candidate is hard to make 
and so far no
polarized target data, which is very helpful, is available.

The many successes of the simple quark model (QM) are somewhat of a 
mystery.  In the QM
all baryons are $qqq$ states and they can be classified in SU(3) 
multiplets.  This
implies one antisymmetric singlet state and two mixed octets with the 
same $J^P$, plus one
related totally symmetric decuplet.  All mesons are grouped in nonets 
with the same $J^P$;
each one consists of one singlet and 8 octet states.  The QM accounts 
for the main
features, though not the details of the masses, widths, decay 
branching rations,
magnetic dipole moments and strength of electromagnetic couplings of 
all  established
baryons and mesons.  There is as yet no compelling evidence for a 
gluon degree of
freedom.  This leads us to the important question: \lq\lq Where in 
hadron physics at
low energy is the glue?"

Recently, a new large facility was put into operation for the important goal of
discovering and subsequently determining the basic properties of the 
quark-gluon
plasma.  This new state of matter is high on the list of interests of 
many a nuclear
theorist.  It is important also to make the necessary investments to enable the
measurements of the properties of hadronic matter at standard density and
temperature.  As a minimum we need to measure the mass, width, and branching
ratios of the excited states of the 6 light baryon and 3 meson families, in
particular of the $\Sigma^*,\ \Xi^*$ and $\Omega^*$ states below 4 GeV.

d)  Broken symmetries play a major role in the theory of the strong 
interactions.
We think here of chiral symmetry, flavor symmetry with its sub areas 
of isospin and
SU(3) symmetry and the U(1) symmetry.  There has been considerable 
speculation on
the occurrence of chiral restoration at intermediate energy [13] and 
some on U(1)
restoration.  Do there exist other symmetries which play a role in 
the structure
of the baryons at ordinary density due to a diquark substructure of 
the baryons [14]?
Are there other clues besides the occurrence of parity doublets?  The 
parity doublets are
seen in the baryon spectra above a certain mass which is 1600 MeV for the $N^*$
family.  By contrast they do not occur for the meson families.  We 
recall the occurrence of
the extraordinary S- wave $\eta$ decays of the low lying $J^P =
  \frac {1}{2} ^-$ octet baryons, the $N$(1535), $\Lambda$(1670) and 
$\Sigma$(1750).  It is
of interest to see if this special $\eta$-decay feature occurs in the 
$\Xi$ family
as well.  The above set of 3 states suggests the existence of a $\Xi$ state
with $J^P = \frac {1}{2} ^-$ and a mass of approximately 1875 MeV.

The first excited state in the $N^*$ family is the $N(1440)\frac 
{1}{2}^+$ and in the
$\Lambda^*$ it is the $\Lambda (1600)\frac {1}{2} ^+$.  They have the 
same spin/parity as
the family's ground state and they are relatively broad.  The same 
holds for the
$\Delta (1600)\frac{3}{2}^+$, which is a decuplet state.  We 
speculate that this
phenomenon applies to the other baryon families as well and urge a search for a
$\Sigma^*\ \frac{1}{2}^+$  octet state of mass $\sim1680$ MeV as well as a
$\Xi ^*\ \frac{1}{2}^+$ octet state around 1800 MeV.  Furthermore we anticipate
a $\Sigma^*\ \frac{3}{2}^+$ decuplet state of mass $\sim$ 1770 MeV, a $\Xi ^*\
\frac{3}{2}^+$ decuplet state around $\sim$ 1900 MeV and a $\Omega  ^*\
\frac{3}{2}^+$ at $\sim$ 2050 MeV. All are expected to be relatively broad
states.

e) Creation of matter out of energy is another important subject.  The
Einstein condition:
\[
E^2 = p^2 c^2 + m^2c^4
\]
which is incorporated in QCD does not say anything about changing energy into
$q\bar q$ pairs and gluons.  It would be quite useful to understand 
how the features
of mass creation such as the simple reaction $\pi^- p\to\pi^0\pi^0 n$
come out of QCD.  Chiral perturbation theory has been very successful in
giving a detailed account of threshold $\pi^0$ photoproduction.  The
next step is the threshold production of the other two Goldstone
bosons, the $K$ and $\eta$.  The creation of new particles at low
energy is related to a process called hadronization in high energy
physics.  This is the creation of hadrons by a high energy quark or
gluon in jet-type events.

It is desirable to make a systematic study of meson and
baryon-antibaryon production using beams of $\pi^\pm, K^\pm, p, \bar
p, \ e^-$ and $\gamma$.

f) The development of effective models for the making of practical
calculations of hadron-hadron scattering, resonance decay, and particle
production is a mandatory aspect of our field. Which are the most
suitable degrees of freedom to use?  How complementary are the
different models and how well does each model implement the symmetries
of QCD?  There is a large variety of strong interaction models to choose
from: chiral perturbation theory, the $N_c^{-1}$ expansion, Skyrmions,
meson exchange, nuclear potentials, vector meson dominance, bag models,
and so forth.

\section{Input parameters to QCD and to effective Lagrangians.}

QCD needs 3 types of input:

1. $\Lambda_{QCD}$, the fundamental strength scale; one can also 
choose the running
coupling constant $\alpha_s$. The experimental determination of this 
input is done using
high energy experiments that operate in the perturbative regime of 
QCD.  The present
precision achieved is several percent.  To go much beyond that one 
must first learn how to
make the hadronic corrections which belong in the domain of 
non-perturbative QCD. The
hadronic corrections are made using a suitable effective Lagrangian. 
Efforts are underway
to use large computers and many new algorithms for lattice-gauge calculations;
they still have a long way to go before becoming practical.

2. The masses of the six quarks.  Since there are no free quarks the mass
determinations are a joint endeavor of theory and experiment. What we know best
  is the ratio of quark mass differences.

3. The $\theta$ parameter of QCD.  The upper limit on the electric 
dipole moment
of the  neutron provides an upper bound $\theta\le 2\times 10^{-10}$.  It
does not play a significant role in QCD at present.

The effective models needed for calculating strong interaction results in the
non-perturbative regime of QCD require a large set of inputs.  Many of these
are discussed in detail at our MENU symposium and related workshops and
conferences such as MESON 2001, Baryon 2001, NSTAR 2001 and PANIC.  We mention
a few inputs of the many needed:
\begin{itemize}
\item a) The meson decay constants: $F_\pi, F_\eta, F_K$, etc.
\item b) The meson-nucleon coupling constants: $G_{\pi N N}$, 
$G_{\eta N N}$, etc.
\item c) The meson-meson scattering lengths: $a_{\pi\pi}$, $a 
_{\eta\pi}$, etc.
\item d) Various mixing angles such as the SU(3) singlet-octet mixing 
angles for
the pseudoscalar and vector mesons etc.
\item e) Hadron form factors for the proton, neutron and the light mesons.
\end{itemize}

\section{Goals of Hadron Physics.}

The meeting entitled \lq\lq Key Issues in Hadronic Physics held in 
Duck, NC on Nov.\ 6--9 2001 provided the following useful formulation
of the overall 
objectives of
our field [1].

   The primary goals of hadronic physics are to determine the relevant
degrees of freedom that govern hadronic phenomena at all scales, to establish
the connection of these degrees of freedom to the parameters and fundamental
fields of QCD, and to use our understanding of QCD to quantitatively describe
a wide array of hadronic phenomena, ranging from terrestrial nuclear physics
to the behavior of matter in the early universe. We list below some very
specific goals for the near future in our area.

1. The determination of the ratio of the difference and sum of the up and down
current-quark masses,
\[
R = \frac{m_d - m_u}{m_d + m_u}.
\]
The ratio eliminates many of the conceptual problems arising from the fact that
free quarks have never been observed, despite much searching, and are not
expected to be seen as free particles. Experimentally, a nice, clean way to
investigate the up-down quark mass difference is by a measurement of 
the absolute
decay rate for the decay $\eta\to 3\pi$. Another good way is by an accurate
measurement of the ratio
\[
\Gamma (\eta\to\pi^+\pi^-\pi^0)/ \Gamma (\eta\to 3\pi^0).
\]
Other ways include 2 special ratios of meson decay modes,
\[
\Gamma  (\eta^{\prime}\to\eta\ 2\pi^0)/\Gamma(\eta^\prime\to 3\pi^0),
  \]
and
\[
\Gamma(\psi^\prime\to\eta\psi)/\Gamma(\psi^\prime\to\pi^0\psi).
\]
2. The determination of the ratio of quark-masses
\[
\frac{m_s - m_u}{m_d + m_u}.
\]
This requires careful measurements of SU(3) flavor breaking such as 
probed in the
Gell-Mann-Okubo octet and Gell-Mann decuplet mass relations.  This 
should be done
for different spin/parity baryons.

3. It would be nice to have a full QCD calculation of the neutron-proton mass
difference which is known experimentally to an accuracy of a few 
parts in $10^7$.

4. It is time that we measured the magnetic dipole moment of other 
particles than
the ground state octet and decuplet baryons.

5. The determination of $\pi^0-\eta$ mixing in different nuclear environments.

6. Accurate measurements of the inputs to the various effective theories such
as the meson-nucleon coupling constants, the meson-meson scattering 
lengths, the
decay constants, and so forth.

\section{The importance of Hadron Physics in testing the Standard Model.}

The Standard Model (SM) of electroweak interactions has been subjected to many
experimental tests and passed them all with flying colors.  Yet, the 
SM is called
a model and not a theory.  It suffers from having 17 input parameters 
which must
be extracted from many experiments
  and this does not include 3 or 6 possible neutrino masses and several mixing
angles.  The SM does not explain why there are three families of fundamental
fermions; parity violation is inserted into the theory by choosing a 
left-handed
doublet and a right-handed singlet of fermions in the input 
structure; the nature
of the spontaneous symmetry breaking that generates the mass of the elementary
ferimons is unknown; these are just some of the shortcomings of the SM.  New
ideas, such as supersymmetry, are going beyond the classical SM. 
Experiments which
explore the limits of the SM,  euphemistically called \lq\lq Searches
for New Physics", are limited by the insufficiently known hadronic corrections.
Since there is no analytic solution to QCD in the vast non-perturbative domain,
one is forced to rely on effective Lagrangians and on models. Below are a few
examples of recent precision measurements in the frontier of the SM, they
involve real as well as virtual hadrons.  The usefulness of the experiments is
limited mainly by the uncertainties in the hadronic corrections.

1. The latest measurement of $g-2$, the anomalous g-value of the muon magnetic
moment, made at BNL, is of sufficient precision to be sensitive to \lq\lq New
Physics" [7].  $g-2$ calculations have 3 components: a. the pure QED 
part, which is
known to 0.025 $ppm$; b. the electroweak part known to 0.03 $ppm$; c. the
hadronic corrections, which have an uncertainty of 0.57 $ppm$, originate in
the higher order electromagnetic interactions of leptons arising from virtual
hadronic contributions to the photon propagator.  An important contribution
comes from light-light scattering where the error is due mainly to the poorly
known form factors of the intermediate $\pi,\eta$ and $\eta^\prime$. 
This can be
improved if new measurements are made of single and particularly of 
double Dalitz
decays such as $\eta\to e^+e^-\mu^+\mu^-$.

2. An important investigation into the nature of $CP$-violation is the
determination of the ratio of the direct to indirect $CP$- violation 
parameters,
$\epsilon$ to $\epsilon^\prime$ [8,9].  This is done by measuring the ratio of
$K^{0}_L$ and $K^{0}_{S}$ decay into $\pi^+\pi^-$ and $\pi^0\pi^0$ at the level
$10^{-3}$ to $10^{-4}$.  It is immediately clear that at this level 
the hadronic
corrections originating in the $s-d$ quark transition and the isospin 
breaking due
to the $u-d$ quark mass difference are major, they greatly cloud the
interpretation of the precision measurements.

3. $CP$ violation has finally been seen outside the $K^0 - \bar{K} ^0$ system
namely in $B$ decays at the $B/\bar{B}$ and Belle storage [15,16] rings.  The
data are subject to similar type corrections as in $K$-decays.

4. The Standard Model requires that the Kobayashi-Maskawa-Cabbibo (CKM) matrix
is a unitary matrix.  This has not been verified to any desirable precision.
The CKM matrix provides a test in which supersymmetry could show up before the
LHC turns on.  This test requires only accurate values for $V_{ud}$ 
and $V_{us}$;
$V_{ub}$ is very small and is known well enough for the purpose.  $V_{ud}$ is
being measured in super-allowed beta decays, also in neutron decay. $V_{us}$
must be obtained from a precision measurement of the decay rate and spectrum of
$K_{e3}$ decay, $K^+\to\pi^0 e^+v$.  The limits are due to the uncertainties in
the handling of SU(3) breaking and isospin violation, if the latter is a
surprise it is due to $\pi^0 -\eta$ mixing in the final state.

5. Another area where the hadronic corrections limit the accuracy of the
results is the determination of the input parameters to QCD and the SM.  The
success of perturbative QCD may be illustrated compactly by the internal
agreement on the average value of the strong coupling parameter
$\alpha_s$ obtained in 12 different ways [5].  The average value 
quoted [5] is
$\alpha_s(M_z) = 0.1181 \pm 0.002$.  To do much better one needs to know the
masses of the quarks for which one depends on non-perturbative QCD!

Our conclusion is that major advances in the frontier of particle and nuclear
physics depend on the advances that are being made in the physics of the
non-perturbative sector.

\begin{center}\bf {Acknowledgements.}
\end{center}

It is my pleasure to express the heartfelt appreciation of the participants for
the dedicated efforts in organizing this very successful 
International Symposium
MENU-IX made by Bill Briscoe and Helmut Haberzetll and their dedicated staff.
Their countless hours of unheralded work on visitor matters, subsidies, program
balance and all the organizational details have paid off handsomely. 
MENU-IX will
be recorded in the annals of hadron-physics history as a four-star conference,
the highest rating we can bestow.


\end{document}